\documentstyle[12pt]{article}
\begin{document}

\title{Natural Doublet-Triplet Splitting in $SU(N) \times U(1)$}

\author{{\bf S.M. Barr} \\
Bartol Research Institute \\ University of Delaware \\
Newark, Delaware 19716}

\date{\today}

\maketitle

\begin{abstract}
It is shown that natural doublet-triplet splitting
can be achieved in a relatively simple way in supersymmetric
$SU(N) \times U(1)$ models with $N>5$.
\end{abstract}

\newpage

Much more attention has been paid to $SO(10)$ unification than to
unification based on $SU(N)$ with $N>5$. Part of the appeal of
$SO(10)$ is that the observed quarks and leptons fit perfectly into
spinor multiplets. As pointed out long ago, however, there is also a
simple explanation of the observed fermion spectrum in the context
of large unitary groups. Any anomaly-free set of antisymmetric
tensor multiplets of fermions in $SU(N)$, with $N>5$, will decompose
under the Standard Model subgroup into families with the observed
quantum numbers \cite{survival}. Interestingly, in such $SU(N)$
models, the different families can come from tensor multiplets of
different size, unlike the situation in $SU(5)$ and $SO(10)$, where
the families arise from having three copies of the same multiplets
($3 \times ( {\bf 10} + \overline{{\bf 5}})$ or $3 \times ({\bf
16})$, respectively). This feature may be relevant to explaining the
observed mass hierarchy among the families \cite{suNflavor}.

One problem with $SU(N)$ unification, however, is achieving
doublet-triplet splitting \cite{splitting} in a way that is both
technically natural and economical. In $SU(5)$, the only technically
natural doublet-triplet splitting is by the missing partner
mechanism \cite{mpm}, but it requires Higgs fields in rather large
tensor multiplets, specifically $H^{[\alpha \beta \gamma]}_{[\delta
\epsilon]} = {\bf 50}$ and $H^{[\alpha \beta]}_{[\delta \epsilon]} =
{\bf 75}$. This can be generalized to $SU(N)$, $N>5$, but has the
same drawback there. The ``sliding singlet" mechanism can be made to
work in $SU(N)$ models with $N>5$ \cite{sliding}, but it requires
the Higgs superpotential to have a very special form
\cite{su6sliding}.  There is also the so-called GIFT mechanism
\cite{gift}, which can
be implemented in models with unitary groups larger than $SU(5)$.

As is well known, the missing partner mechanism can be implemented
in an extremely economical way in supersymmetric ``flipped $SU(5)$"
\cite{mpmFlipped} (the group of which is really $SU(5) \times U(1)$).
In this paper, we show that this can be
generalized to supersymmetric $SU(N) \times U(1)$.

The possibility of a ``flipped" breaking of $SU(5) \times U(1)$ to
the Standard Model in addition to the ``Georgi-Glashow" breaking can
be understood group-theoretically as arising from the fact that
$SU(5) \times U(1)$ is embeddable in $SO(10)$ in two distinct ways.
Even if the $SU(5) \times U(1)$ gauge group is not actually embedded
in an $SO(10)$ gauge group, the flipped breaking requires that the
$U(1)$ charges be ``$SO(10)$-like", i.e. that the fermion multiplets
${\bf 10}$, $\overline{{\bf 5}}$, and ${\bf 1}$ have $U(1)$ charges
1, -3, and 5, just as though they came from a spinor of $SO(10)$,
and similarly for the charges of Higgs multiplets.

In a realistic $SU(N) \times U(1)$ model, with $N>5$, however, the
fermion multiplets are not embeddable in an $SO(10)$ (nor, in most
cases, in any larger orthogonal group). Nevertheless, in many simple
$SU(N) \times U(1)$ models it happens that the $U(1)$ charges of the
quark/lepton multiplets are forced to be ``$SO(10)$-like" by anomaly
cancellation. That is enough to allow for flipped breaking of $SU(N)
\times U(1)$ down to the Standard Model.  With such breaking, as
will be seen, an economical implementation of the missing partner
mechanism is possible. Unlike the case of $SU(5) \times U(1)$,
however, it seems to require that some of the known quarks and
leptons obtain mass from higher-dimension effective Yukawa
operators.

We shall briefly review how the missing partner mechanism works in
flipped $SU(5) \times U(1)$. Then we shall show explicitly how it
can be generalized to $SU(6) \times U(1)$ and $SU(7) \times U(1)$
models. The further generalization to larger unitary groups is
relatively straightforward.

In the simplest supersymmetric flipped $SU(5)$ model, the quarks and
leptons of a family are in ${\bf 10}^1 + \overline{{\bf 5}}^{-3} +
{\bf 1}^5$ of $SU(5) \times U(1)_X$ and the Higgs multiplets include
${\bf 10}_H^1 + \overline{{\bf 10}}_H^{\;-1} + {\bf 5}_H^{-2} +
\overline{{\bf 5}}_H^2$. The breaking of $SU(5) \times U(1)_X
\rightarrow SU(3)_c \times SU(2)_c \times U(1)_Y$ is done by
$\langle H^{12} \rangle \in {\bf 10}_H^1$ and $\langle H_{12}
\rangle \in \overline{{\bf 10}}_H^{-1}$. This leaves unbroken $Y/2 =
\frac{1}{5}(-Y_5/2 + X)$, where $Y_5/2$ is the $SU(5)$ generator
$diag(\frac{1}{2}, \frac{1}{2},
-\frac{1}{3},-\frac{1}{3},-\frac{1}{3})$. The Higgs superpotential
is assumed to have the couplings ${\bf 10}_H^1 {\bf 10}_H^1 {\bf
5}_H^{-2} + \overline{{\bf 10}}_H^{-1} \overline{{\bf 10}}_H^{-1}
\overline{{\bf 5}}_H^{2}$, but not the coupling ${\bf 5}_H^{-2}
\overline{{\bf 5}}_H^2$. Then the superheavy masses of the
Higgs(ino) doublets and triplets can be represented as follows:

\begin{equation}
\begin{array}{ccccccc} {\bf 5}_H^{-2} & \;\;\; & {\bf 10}_H^1 & \;\;\; &
\overline{{\bf 10}}_H^{-1} & \;\;\; & \overline{{\bf 5}}_H^2 \\
\hline
 & & \langle H^{12} \rangle & & \langle H_{12} \rangle & & \\
H^i & & \left[ H^{ia} \right] & & \left[ H_{ia} \right] & & H_i \\
H^a & \longleftrightarrow & \epsilon_{abc} H^{bc} & &
\epsilon^{abc} H_{bc} & \longleftrightarrow & H_a \end{array}
\end{equation}
\begin{center}
{\bf Table I.}
\end{center}

\noindent where $i,j = 1,2$ are $SU(2)_L$ indices, and $a,b,c =
3,4,5$ are $SU(3)_c$ indices. The arrows connecting two multiplets
represent superheavy Dirac masses, in this case produced by the
couplings ${\bf 10}_H^1 {\bf 10}_H^1 {\bf 5}_H^{-2} + \overline{{\bf
10}}_H^{-1} \overline{{\bf 10}}_H^{-1} \overline{{\bf 5}}_H^{2}$,
which contain $\epsilon_{12abc} \langle H^{12} \rangle H^{bc} H^a  +
\epsilon^{12abc} \langle H_{12} \rangle H_{bc} H_a $. The square
brackets around $H^{ia}$ and $H_{ia}$ in Table I represent the fact
that these multiplets are eaten by the Higgs mechanism when $H^{12}$
and $H_{12}$ obtain vacuum expectation values (VEVs). One sees that
the doublet fields $H^i$ and $H_i$ do not acquire mass as they have
no weak-doublet, color-singlet ``partners" in the ${\bf 10}_H +
\overline{{\bf 10}}_H$. These are the doublets of the MSSM and
couple to the quark and lepton multiplets through the usual terms
$({\bf 10}^1 {\bf 10}^1){\bf 5}_H^{-2} + ({\bf 10}^1 \overline{{\bf
5}}^{\;-3}) \overline{{\bf 5}}_H^2  + (\overline{{\bf 5}}^{\;-3}
{\bf 1}^5) {\bf 5}_H^{-2}$.

Let us now see how this mechanism can be generalized to the gauge
groups $SU(6) \times U(1)$. The details which we now present are
more involved than in flipped $SU(5)$, simply because the group has
higher rank and there is another step of symmetry breaking required,
and also because the multiplets of $SU(6)$ contain more Standard
Model fields, which have to be kept track of. This will be even more
the case when we turn to $SU(7) \times U(1)$. But the essence of
what is going on is the same as in flipped $SU(5)$, and is basically
simple: the missing-partner mechanism of flipped $SU(5)$ is being
embedded in larger unitary groups. The matter (quark and lepton)
content in each case is essentially minimal, and the structure of
the Higgs superpotentials (in Eqs. (2) and (4)) is simple. We now go
through the details carefully to demonstrate that, indeed, the
mechanism can be implented in a realistic way and no obstruction
arises for these larger groups.

A family in $SU(6)$ consists of
$\psi^{AB} + \psi_A + \psi_A = {\bf 15} + \overline{{\bf 6}} +
\overline{{\bf 6}}$. (Capital letters $A, B = 1,2, ..., 6$ denote
$SU(6)$ indices.) Up to an overall normalization, there is only one
way to assign $U(1)$ charges to the quarks and leptons in a
family-independent way that is consistent with anomaly cancellation,
namely ${\bf 15}^0 + \overline{{\bf 6}}^1 + \overline{{\bf
6}}^{\;-1}$. (The contribution to the anomalies of the Higgsinos
would vanish, if the Higgs multiplets are in conjugate pairs under
the gauge group as would generally be required to avoid D-term
breaking of supersymmetry.) The $U(1)$ charges of the Higgs
superfields are determined by their Yukawa couplings to the quark
and lepton supermultiplets. Higgs superfields are needed in the
following multiplets: ${\bf 15}_H^0 + \overline{{\bf 6}}_H^1 +
\overline{{\bf 6}}_H^{\;-1}$ plus their conjugate multiplets.

The breaking to the Standard Model group is done by the vacuum
expectation values $\langle H^{(-1)}_6 \rangle \in \overline{{\bf
6}}_H^{-1}$ and $\langle H^{12} \rangle \in {\bf 15}_H^0$ and the
corresponding vacuum expectation values in the conjugate Higgs
multiplets. (We use the superscripts $(\pm 1)$ to distinguish the
two anti-fundamental Higgs multiplets based on their $U(1)$ charge.)
The first of these VEVs breaks $SU(6) \times U(1) \longrightarrow
SU(5) \times U(1)_X$. If $T_6$ is the $SU(6)$ generator
$diag(1,1,1,1,1,-5)$ and $T_1$ is the generator of the original
$U(1)$ as normalized above, then the generator of the unbroken
$U(1)_X$ is given by $X = \frac{1}{2} (T_6 + 5 T_1)$. This is clear,
since $H^{(-1)}_6$ has $T_6 = 5$ and $T_1 = -1$. The field $H^{12}
\in {\bf 15}_H^0$ then does the usual ``flipped" breaking of $SU(5)
\times U(1)_X$ down to the Standard Model group.  One can see that
it has the correct $SU(5) \times (1)_X$ charges to do this. It has
$T_6 =2$, $T_1 =0$, and thus $X = 1$. It also has $Y_5/2 = 1$, where
$Y_5/2$ is the $SU(6)$ generator $diag(\frac{1}{2}, \frac{1}{2},
-\frac{1}{3}, -\frac{1}{3}, -\frac{1}{3}, 0)$, which is also a
generator of the $SU(5)$ subgroup. Therefore, its VEV breaks $SU(5)
\times U(1)_X$ to the Standard Model group, with $Y/2 =
\frac{1}{5}(-Y_5/2 + X) = \frac{1}{5}(-Y_5/2 + \frac{1}{2} T_6 +
\frac{5}{2} T_1)$.

The decomposition of the quarks-lepton and Higgs multiplets is given
in Table II.

\begin{displaymath}
\begin{array}{ccccccc}
SU(6)^{T_1} & \;\;\;\;\; & SU(5)^{T_6,T_1} & \;\;\;\;\; & SU(5)^X &
\;\;\;\; & \\
\hline \psi^{AB} = {\bf 15}^0 & & \psi^{\alpha \beta} = {\bf
10}^{2,0} & & {\bf 10}^1 & & light \\
& & \psi^{\alpha 6} = {\bf 5}^{-4,0} & & {\bf 5}^{-2} & & SUPERHEAVY \\
\hline \psi^{(1)}_A = \overline{{\bf 6}}^1 & & \psi^{(1)}_{\alpha}=
\overline{{\bf 5}}^{\;-1,1} & &
\overline{{\bf 5}}^2 & & SUPERHEAVY \\
& & \psi^{(1)}_6 = {\bf 1}^{5,1} & & {\bf 1}^5 & & light \\
\hline \psi^{(-1)}_A = \overline{{\bf 6}}^{\; -1} & &
\psi^{(-1)}_{\alpha}= \overline{{\bf 5}}^{\;-1,-1} & &
\overline{{\bf 5}}^{-3} & & light \\
& & \psi^{(-1)}_6 = {\bf 1}^{5,-1} & & {\bf 1}^0 & & SUPERHEAVY \\
\hline H^{AB} = {\bf 15}_H^0 & & H^{\alpha \beta} =
{\bf 10}_H^{2,0} & & {\bf 10}_H^1 & & \langle H^{12} \rangle \sim M_G \\
& & H^{\alpha 6} = {\bf 5}_H^{-4,0} & & {\bf 5}_H^{-2} & &
\langle H^{26} \rangle \sim M_W \\
\hline H^{(1)}_A = \overline{{\bf 6}}_H^1 & & H^{(1)}_{\alpha}=
\overline{{\bf 5}}_H^{\;-1,1} & &
\overline{{\bf 5}}_H^2 & &  \\
& & H^{(1)}_6 = {\bf 1}_H^{5,1} & & {\bf 1}_H^5 & &  \\
\hline H^{(-1)}_A = \overline{{\bf 6}}_H^{\; -1} & &
H^{(-1)}_{\alpha}= \overline{{\bf 5}}_H^{\;-1,-1} & &
\overline{{\bf 5}}_H^{-3} & &  \langle H_2^{(-1)} \rangle \sim M_W \\
& & H^{(-1)}_6 = {\bf 1}_H^{5,-1} & & {\bf 1}_H^0 & & \langle
H^{(-1)}_6
\rangle \sim M_G \\
\hline
\end{array}
\end{displaymath}

\begin{center}
{\bf Table II.}
\end{center}

\noindent Note that the $X$ charges are exactly those that would
result if $SU(5) \times U(1)_X$ were embedded in an $SO(10)$, even
though obviously $SU(6) \times U(1)$ is not a subgroup of $SO(10)$.
This can be understood in two ways: (1) in terms of the structure of
$E_6$, or (2) as the result of anomaly cancellation. The argument
based on $E_6$ is as follows. $E_6$ contains the subgroups $E_6
\supset SU(6) \times SU(2) \supset SU(6) \times U(1)$ under which
${\bf 27} \longrightarrow ({\bf 15}, {\bf 1}) + (\overline{{\bf 6}},
{\bf 2}) \longrightarrow {\bf 15}^0 + \overline{{\bf 6}}^1 +
\overline{{\bf 6}}^{\;-1}$. Note that these $SU(6) \times U(1)$
charges assignments are just those assumed in Table II. But also
$E_6$ contains the subgroups $E_6 \supset SO(10) \supset SU(5)
\times U(1)_X$. The fact that the flipped $SU(5) \times U(1)_X$ is a
subgroup of the $SU(6) \times U(1)$ in Table II can therefore be
understood as a consequence of the group theory of $E_6$. This
explanation is special to $SU(6) \times U(1)$, however, and does not
generalize to $SU(N) \times U(1)$ with $N>6$.

The anomaly cancellation argument is more general. The point is that
if $SU(N) \times U(1)$ is broken to $SU(5) \times U(1)_X$ and the
light families end up as ${\bf 10} + \overline{{\bf 5}} + {\bf 1}$
then $X$ has to have the ``$SO(10)$-like" assignments ${\bf 10}^1 +
\overline{{\bf 5}}^{\;-3} + {\bf 1}^5$ because of anomaly freedom.
(This is so, even if each light family contains several extra
$SU(5)$ singlets.) This is the crucial point that makes it possible
to have ``flipped" breaking and to implement the missing partner
mechanism in an economical way in $SU(N) \times U(1)$ models, even
if they have particle content that cannot be understood in terms of
orthogonal or exceptional groups.

In the $SU(6) \times U(1)$ model, the ``flipped" breaking of the
$SU(5) \times U(1)_X$ subgroup down to the Standard Model group is
done (as in ordinary flipped $SU(5)$) by $\langle H^{12} \rangle$
and $ \langle H_{12} \rangle$. As shown in Table II, these are
contained in the ${\bf 15}^0 + \overline{{\bf 15}}^0$ of $SU(6)
\times U(1)$. The doublet-triplet splitting can be achieved through
a Higgs superpotential that contains terms of the following form:

\begin{equation}
\begin{array}{ll}
W_H \supset & {\bf 15}_H^0 {\bf 15}_H^0 {\bf 15}_H^0 +
\overline{{\bf 15}}_H^0 \overline{{\bf 15}}_H^0 \overline{{\bf
15}}_H^0 \\ & + M_1(\overline{{\bf 6}}_H^1 {\bf 6}_H^{-1}) +
(\overline{{\bf 6}}_H^{-1} {\bf 6}_H^1 - M_2^2)Z  + (\overline{{\bf
15}}_H^0 {\bf 15}_H^0 - M_3^2)Z', \end{array}
\end{equation}

\noindent where $Z$ and $Z'$ are gauge singlets that are integrated
out to give nonzero superlarge VEVs to the components that break
$SU(6) \times U(1)$ to the Standard Models group at superlarge
scales. There may be other terms in the Higgs superpotential as
well. We do not write terms of the form $M(\overline{{\bf 15}}_H^0
{\bf 15}_H^0)$ and $M(\overline{{\bf 6}}_H^{\;-1} {\bf 6}_H^1)$ in
Eq. (2), since such terms can be eliminated simply by shifting the
singlet superfields $Z$ and $Z'$. (Or one could forbid such terms
with a discrete symmetry.)

The color-triplet Higgs(ino) fields in ${\bf 15}_H^0$ and
$\overline{{\bf 15}}_H^0$ obtain mass from the first terms in Eq.
(2), which contain $\epsilon_{12abc6} \langle H^{12} \rangle H^{bc}
H^{a6} + \epsilon^{12abc6} \langle H_{12} \rangle H_{bc} H_{a6}$.
The doublet-triplet splitting works essentially as shown in Eq. (1).
There are altogether three triplet Higgs multiplets with Standard
Model quantum numbers $(3,1,-\frac{1}{3})$, which are $H^{a6} \in
{\bf 15}_H^0$, $H_{bc} \in \overline{{\bf 15}}_H^0$, and $H^{(-1)a}
\in {\bf 6}_H^{\;-1}$; and there are three anti-triplet Higgs in
$(\overline{3}, 1, +\frac{1}{3})$ that are conjugate to these. The
mass matrix of these colored Higgs(ino) fields with $Y/2 = \pm
\frac{1}{3}$ is

\begin{equation}
W_{(3,1,-1/3)}= \left( H^{a6}, \epsilon^{abc} H_{bc}, H^{(-1)a}
\right) \left(
\begin{array}{ccc}
0 & \langle H^{12} \rangle & 0 \\
\langle H_{12} \rangle & 0 & 0 \\ 0 & 0 & M_1
\end{array} \right) \left( \begin{array}{c}
H_{a6} \\ \epsilon_{abc} H^{bc} \\ H^{(1)}_a \end{array} \right).
\end{equation}

\noindent All these triplets acquire superlarge mass. There are also
a pair of triplets with Standard Model quantum numbers $(3, 1,
\frac{2}{3}) + (\overline{3}, 1, -\frac{2}{3})$, namely $H^{(1)a}
\in {\bf 6}_H^1$ and $H^{(-1)}_a \in \overline{{\bf 6}}_H^{\;-1}$.
These obtain no mass from the terms in Eq. (2), and indeed are
goldstone fields that get eaten by the Higgs mechanism when
$H^{(1)6}$ and $H^{(-1)}_6$ get superlarge VEVs.

There are altogether three doublet Higgs multiplets with Standard
Model quantum numbers $(1,2, \frac{1}{2})$, which are $H_{i6} \in
\overline{{\bf 15}}_H^0$, $H^{(1)i} \in {\bf 6}_H^1$, and $H^{(1)}_i
\in \overline{{\bf 6}}_H^1$; and there are three doublet Higgs with
$(1, 2, -\frac{1}{2})$ that are conjugate representations to these.
The mass matrix of these doublet Higgs is

\begin{equation}
W_{(1,2,1/2)}= \left( H_{i6}, H^{(1)i}, H^{(1)}_i \right) \left(
\begin{array}{ccc}
0 & 0 & 0 \\
0 & 0 & 0 \\ 0 & 0 & M_1
\end{array} \right) \left( \begin{array}{c}
H^{i6} \\ H^{(-1)}_i \\ H^{(-1)i} \end{array} \right).
\end{equation}

\noindent Two pairs of doublet Higgs fields are left massless by the
mass matrix in Eq. (4). (This is due to the absence of the terms of
the form $M(\overline{{\bf 15}}_H^0 {\bf 15}_H^0)$ and
$M(\overline{{\bf 6}}_H^{\;-1} {\bf 6}_H^1)$ in Eq. (2), which was
discussed above. That is to say, it is due to the fact that $\langle
Z \rangle =0$ and $\langle Z' \rangle = 0$.) Of these two massless
pairs of doublets, one pair consists of goldstone fields that get
eaten by the Higgs mechanism, and the other pair are the two doublet
Higgs multiplets of the MSSM. The MSSM doublet with
$(1,2,\frac{1}{2})$ is a linear combination of the doublets in
$\overline{{\bf 15}}_H^0$ and ${\bf 6}_H^1$, and the $(1,2,-
\frac{1}{2})$ doublet of the MSSM is a linear combination of the
doublets in ${\bf 15}_H^0$ and $\overline{{\bf 6}}_H^{\;-1}$.
Finally, there are also the goldstone modes $H^{ia} \in {\bf
15}_H^0$ and $H_{ia} \in \overline{{\bf 15}}_H^0$ that are eaten by
the Higgs mechanism. The absence of uneaten goldstone Higgs fields
implies that this minimum is not continuously connected to other
degenerate minima, which is a condition for satisfactory breaking of
$SU(6) \times U(1)$ to the Standard Model.

Although we arrived at this set-up by generalizing flipped $SU(5)
\times U(1)$ to $SU(6) \times U(1)$, it may not be obvious from
examining the mass matrices in Eqs. (3) and (4) that the
doublet-triplet splitting is being accomplished here by the missing
partner mechanism. It can be seen more readily if we note that the
${\bf 15}_H^0$ contains both a $(3,1,-\frac{1}{3})$ (namely
$H^{a6}$) and a $(\overline{3}, 1, +\frac{1}{3})$ (namely
$\epsilon_{abc} H^{bc}$), and these are ``partners" of each other in
a mass term of the form $\epsilon_{12abc6} H^{a6} H^{bc} \langle
H^{12} \rangle$. On the other hand, in the ${\bf 15}^0_H$ there is a
$(1,2,-\frac{1}{2})$ (namely $H^{i6}$), but no $(1,2,+\frac{1}{2})$.
That is, the ``partner" of the doublet is ``missing". The same is
true for the $\overline{{\bf 15}}^0_H$. It is important that the
light doublets in ${\bf 15}^0_H$ and $\overline{{\bf 15}}_H^0$ are
not connected to each other by an explicit mass term $M({\bf 15}^0_H
\overline{{\bf 15}}_H^0)$.  This is analogous to the condition in
flipped $SU(5)$ that there is no term $M({\bf 10}_H \overline{{\bf
10}}_H)$. The ${\bf 15}^0_H$ and $\overline{{\bf 15}}_H^0$  of
$SU(6) \times U(1)$ contain the Higgs multiplets that do the
doublet-triplet splitting in flipped $SU(5) \times U(1)$, namely
${\bf 10}_H$, ${\bf 5}_H$, $\overline{{\bf 10}}_H$, and
$\overline{{\bf 5}}_H$.  The ${\bf 6}_H$ and $\overline{{\bf 6}}_H$
multiplets are required to do the breaking of $SU(6) \times U(1)$ to
$SU(5) \times U(1)$. These additional Higgs multiplets slightly
complicate matters, but do not affect the basic missing partner
mechanism. The ${\bf 6}^{(-1)}_H$ and $\overline{{\bf 6}}^{(1)}_H$
are connected by a mass term, and thereby provide ``partners" for
each other. The ${\bf 6}^{(1)}_H$ and $\overline{{\bf 6}}^{(-1)}_H$
are {\it not} connected to each other by a mass term (because
$\langle Z \rangle = 0$), but that does not matter: the triplets in
them are eaten and the extra pair of doublets in them is needed,
since a pair of doublets must be eaten by the Higgs mechanism when
$SU(6)$ breaks to $SU(5)$.

Turning now to the spectrum of quarks and leptons, the extra
vectorlike quarks and leptons in ${\bf 5}^{-2} + \overline{{\bf
5}}^2$ (called ``heavy" in Table I) obtain masses from the term
$({\bf 15}^0 \overline{{\bf 6}}^{1}) \overline{{\bf 6}}_H^{\;-1}$,
which (in $SU(5) \times U(1)_X$ language) contains $({\bf 5}^{-2}
\overline{{\bf 5}}^2) \langle {\bf 1}_H^0 \rangle $. The masses of
the light quarks and leptons arise from the following terms:

\begin{equation}
\begin{array}{ll}
M_D:  ({\bf 15}^0 {\bf 15}^0) {\bf 15}_H^0  & \supset ({\bf 10}^1
{\bf 10}^1) \langle {\bf 5}_H^{-2} \rangle \\
\supset \epsilon_{12abc6} (\psi^{1a} \psi^{bc})\langle H^{26}
\rangle & \supset \epsilon_{12abc} (\psi^{1a}
\psi^{bc}) \langle H^2 \rangle, \\
& \\
M_L:  (\overline{{\bf 6}}^1 \overline{{\bf 6}}^{\;-1}) {\bf 15}_H^0
& \supset ({\bf 1}^5 \overline{{\bf 5}}^{\;-3}) \langle {\bf
5}_H^{-2} \rangle \\ \supset (\psi^{(1)}_6 \psi^{(-1)}_2)\langle
H^{26} \rangle & \supset (\psi
\psi_2) \langle H^2 \rangle, \\
& \\
M_U, M_N: ({\bf 15}^0 \overline{{\bf 6}}^{\;-1}) \overline{{\bf
15}}_H^0 {\bf 6}_H^1/M & \supset ({\bf 10}^1 \overline{{\bf
5}}^{\;-3}) \langle \overline{{\bf 5}}_H^2 \rangle \langle {\bf
1}_H^0 \rangle /M \\ \supset (\psi^{2a} \psi^{(-1)}_a + \psi^{21}
\psi^{(-1)}_1 ) H_{26} H^{(1)6}/M & \supset (\psi^{2a} \psi_a +
\psi^{21} \psi_1) \langle H_2 \rangle \langle H \rangle/M,
\end{array}
\end{equation}

\noindent
where $M_D$, $M_L$, $M_U$, and $M_N$ are the mass matrices
of the down quarks, charged leptons, up quarks, and neutrino Dirac
mass matrix, respectively. Note that unlike the usual flipped
$SU(5)$ models, the mass matrix of the up quarks and the Dirac mass
matrix of the neutrinos must come from a higher-dimension operator.
(The reason is that the light Higgs doublets that break the weak
interactions are purely in $\overline{{\bf 15}}_H^0$ and ${\bf
6}_H^1$ and not in $\overline{{\bf 6}}_H^1$, so that the
renormalizable Yukawa term $({\bf 15}^0 \overline{{\bf 6}}^{\;-1})
\overline{{\bf 6}}_H^1$ does not give mass to these fermions.) This
higher-dimension operator can be obtained by integrating out an
adjoint of superheavy quarks and leptons.

We now turn to the group $SU(7) \times U(1)$. There are several
anomaly-free sets of $SU(7)$ multiplets of fermions that lead to
three families at low energies. We will consider ${\bf 21} +
\overline{{\bf 7}}+ \overline{{\bf 7}}+ \overline{{\bf 7}}$. Other
cases work out in similar ways. There is only one solution for the
$U(1)$ charge assignments (up to an arbitrary normalization) that is
anomaly-free and family-independent, namely ${\bf 21}^0 +
\overline{{\bf 7}}^1 + \overline{{\bf 7}}^{\;-1} + \overline{{\bf
7}}^0$. The simplest set of Higgs supermultiplets to break to the
Standard Model and give mass to the quarks and leptons is $({\bf
35}_H^0 + \overline{{\bf 35}}_H^0)$$+ ({\bf 21}_H^0 + \overline{{\bf
21}}_H^0)$$ + (\overline{{\bf 7}}_H^0 + {\bf 7}_H^0)$$ +
(\overline{{\bf 7}}_H^1 + {\bf 7}_H^{\;-1})$$ + (\overline{{\bf
7}}_H^{\;-1} + {\bf 7}_H^1)$.

The breaking to the Standard Model group can be done by the
following Higgs vacuum expectation values: $\langle H^{(-1)}_6
\rangle \in \overline{{\bf 7}}_H^{\;-1}$, $\langle H^{(0)}_7 \rangle
\in \overline{{\bf 7}}_H^{0}$, $\langle H^{12} \rangle \in {\bf
21}_H^0$, and their conjugates $\langle H^{(1)6} \rangle \in {\bf
7}_H^{1}$, $\langle H^{(0)7} \rangle \in {\bf 7}_H^{0}$, $\langle
H_{12} \rangle \in \overline{{\bf 21}}_H^0$. Denote by $T_1$ the
generator of the $U(1)$, and by $T_7$ the $SU(7)$ generator
$diag(1,1,1,1,1,-5,0)$. The VEVs $\langle H^{(0)}_7 \rangle$ and
$\langle H^{(-1)}_6 \rangle$ break $SU(7) \times U(1)$ down to
$SU(5) \times U(1)_X$, where $X = \frac{1}{2} (T_7 + 5 T_1)$. The
VEVs $\langle H^{12} \rangle$ and $\langle H_{12} \rangle$ do the
``flipped" breaking, leaving unbroken $Y/2 = \frac{1}{5}(-Y_5/2 +
X)$.

The quark-lepton and Higgs multiplets decompose under subgroups of
$SU(7) \times U(1)$ as shown in Table III.

\begin{displaymath}
\begin{array}{ccccccccc}
SU(7)^{T_1} & \; & SU(5)^{T_7,T_1} & \; & SU(5)^X &
\; & quarks/leptons & \; & Higgs \; fields \\
\hline {\bf 21}^0 & & {\bf 10}^{2,0} & & {\bf 10}^1 & & \psi^{\alpha
\beta} = light & & \langle
H^{12} \rangle \sim M_G  \\
& & {\bf 5}^{-4,0} & & {\bf 5}^{-2} & & \psi^{\alpha 6} = heavy & &
\langle H^{26} \rangle \sim M_W \\
& & {\bf 5}^{1,0} & & {\bf 5}^{1/2} & & \psi^{\alpha 7} = heavy & &
\langle H^{\alpha 7} \rangle = 0 \\
& & {\bf 1}^{-5,0} & & {\bf 1}^{-5/2} & & \psi^{67} = heavy & &
\langle H^{67} \rangle = 0 \\
\hline \overline{{\bf 7}}^0 & & \overline{{\bf 5}}^{\;-1,0} & &
\overline{{\bf 5}}^{-1/2} & & \psi^{(0)}_{\alpha} = heavy & &
\langle H^{(0)}_{\alpha} \rangle = 0 \\
& & {\bf 1}^{5,0} & & {\bf 1}^{5/2} & & \psi^{(0)}_6 = heavy & &
\langle H^{(0)}_6 \rangle = 0 \\
& & {\bf 1}^{0,0} & & {\bf 1}^0 & & \psi^{(0)}_7 = heavy & &
\langle H^{(0)}_7 \rangle \sim M_G \\
\hline \overline{{\bf 7}}^1 & & \overline{{\bf 5}}^{\;-1,1} & &
\overline{{\bf 5}}^2 & & \psi^{(1)}_{\alpha} = heavy & &
\langle H^{(1)}_{\alpha} \rangle = 0 \\
& & {\bf 1}^{5,1} & & {\bf 1}^5 & & \psi^{(0)}_6 = light & &
\langle H^{(1)}_6 \rangle = 0 \\
& & {\bf 1}^{0,1} & & {\bf 1}^{5/2} & & \psi^{(0)}_7 = heavy & &
\langle H^{(1)}_7 \rangle = 0 \\
\hline \overline{{\bf 7}}^{\;-1} & & \overline{{\bf 5}}^{\;-1,-1} &
& \overline{{\bf 5}}^{\; -3} & & \psi^{(-1)}_{\alpha} = light & &
\langle H^{(-1)}_{\alpha} \rangle = 0 \\
& & {\bf 1}^{5,-1} & & {\bf 1}^0 & & \psi^{(-1)}_6 = heavy & &
\langle H^{(-1)}_6 \rangle \sim M_G \\
& & {\bf 1}^{0,-1} & & {\bf 1}^{-5/2} & & \psi^{(-1)}_7 = heavy & &
\langle H^{(-1)}_7 \rangle = 0 \\
\hline {\bf 35}^0 & & \overline{{\bf 10}}^{3,0} & & \overline{{\bf
10}}^{3/2} & &   & & \langle
H^{\alpha \beta \gamma} \rangle =0  \\
& & {\bf 10}^{-3,0} & & {\bf 10}^{-3/2} & &  & &
\langle H^{\alpha \beta 6} \rangle = 0 \\
& & {\bf 10}^{2,0} & & {\bf 10}^1 & &  & &
\langle H^{\alpha \beta 7} \rangle = 0 \\
& & {\bf 5}^{-4,0} & & {\bf 5}^{-2} & &  & &
\langle H^{\alpha 67} \rangle = 0 \\
\hline
\end{array}
\end{displaymath}

\begin{center}
{\bf Table III.}
\end{center}

\noindent  The doublets $H_{i6}$ and $H_{i6}$ are the light doublets
of the MSSM and obtain weak-scale VEVs. How they remain light will
be seen next.

The doublet-triplet splitting can be achieved through a Higgs
superpotential that contains terms of the following form:

\begin{equation}
\begin{array}{cl}
W_H \supset & {\bf 35}_H^0 {\bf 21}_H^0 {\bf 21}_H^0 +
\overline{{\bf 35}}_H^0 \overline{{\bf 21}}_H^0 \overline{{\bf
21}}_H^0 + {\bf 21}_H^0 \overline{{\bf 7}}_H^0 \overline{{\bf
7}}_H^{0 \prime}  + \overline{{\bf 21}}_H^0 {\bf 7}_H^0 {\bf 7}_H^{0 \prime}
\\ & \\
&
 + M_1(\overline{{\bf 7}}_H^1 {\bf 7}_H^{-1}) +
(\overline{{\bf 7}}_H^{-1} {\bf 7}_H^1- M_2^2)Z \\ & \\
& + (\overline{{\bf 21}}_H^0 {\bf 21}_H^0 - M_3^2)Z' + M_4
(\overline{{\bf 35}}_H^0 {\bf 35}_H^0) + (\overline{{\bf 7}}_H^0
{\bf 7}_H^0 - M_5^2)Z^{\prime \prime}.
\end{array}
\end{equation}

\noindent There may be other terms in the Higgs superpotential as
well. We do not include in Eq. (6) terms of the form
$M(\overline{{\bf 21}}_H^0 {\bf 21}_H^0)$ and $M(\overline{{\bf
7}}_H^{\;-1} {\bf 7}_H^1)$, as they can be absorbed by a shift in
the singlet superfields $Z$ and $Z'$. Note that there has to be more
than one pair of Higgs fields of type ${\bf 7}^0_H + \overline{{\bf
7}}^0_H$ to have the couplings ${\bf 21}_H^0 \overline{{\bf 7}}_H^0
\overline{{\bf 7}}_H^{0 \prime}  + \overline{{\bf 21}}_H^0 {\bf
7}_H^0 {\bf 7}_H^{0 \prime}$, since the ${\bf 21}$ and
$\overline{{\bf 21}}$ are antisymmetric tensors. We assume that
there are exactly two ${\bf 7}^0_H$, which we distinguish as
unprimed and primed, and similarly for their conjugates.

This is just the straightforward generalization of the $SU(6) \times
U(1)$ case. It is easily shown that all of the doublet and triplet
Higgs fields are either given superheavy mass by the terms in Eq.
(6) or get eaten by the Higgs mechanism, except for a pair of light
doublets, which play the role of the Higgs doublets of the MSSM. To
see this we must first consider what fields get eaten.  The gauge
fields in the coset $SU(7)/SU(5)$ that gets broken include $A^i_6$,
$A^i_7$, $A^a_6$, and $A^a_7$, which have Standard Model quantum
numbers $(1,2,\frac{1}{2})$, $(1,2,0)$, $(3,1,\frac{2}{3})$, and
$(3,1,\frac{1}{6})$, respectively, as well as their conjugates.
There must exist goldstone modes in these representations to get
eaten.

The mass matrix of the colored Higgs(ino) fields that are in $(3,1,
-\frac{1}{3}) + conj.$ is

\begin{equation}
W_{(3,1,-1/3)}= \left( \begin{array}{c} H^{a67} \\
\epsilon^{abc} H_{bc7} \\ H^{a6} \\ \epsilon^{abc} H_{bc}
\\ H^{(-1)a} \end{array} \right)^T \left(
\begin{array}{ccccc}
M_4 & 0 & 0 & \langle H^{12} \rangle & 0 \\
0 & M_4 & \langle H_{12} \rangle & 0 & 0 \\
0 & \langle H^{12} \rangle & 0 & 0 & 0 \\
\langle H_{12} \rangle & 0 & 0 & 0 & 0 \\
0 & 0 & 0 & 0 & M_1
\end{array} \right) \left( \begin{array}{c}
H_{a67} \\ \epsilon_{abc} H^{bc7} \\ H_{a6} \\ \epsilon_{abc} H^{bc} \\
H^{(1)}_a
\end{array} \right).
\end{equation}

\noindent Note that this has no zero eigenvalues, which is
consistent with the fact that no fields of this type eaten. All
these fields get superheavy masses.

The mass matrix of the colored Higgs(ino) fields that are in $(3,1,
\frac{1}{6}) + conj.$ is

\begin{equation}
W_{(3,1,1/6)}= \left( \begin{array}{c} \epsilon^{abc} H_{bc6} \\
H^{a7} \\ H^{(0)a} \\ H^{(0)'a}
\end{array} \right)^T \left(
\begin{array}{cccc}
M_4 & \langle H_{12} \rangle & 0 & 0 \\
\langle H^{12} \rangle & 0 & \langle H^{(0)'}_7 \rangle & \langle H^{(0)}_7 \rangle \\
0 & \langle H^{(0)\prime 7} \rangle & 0 & 0  \\
0 & \langle H^{(0)7} \rangle & 0 & 0  \\
\end{array} \right) \left( \begin{array}{c}
\epsilon_{ab'c'} H^{b'c'6} \\ H_{a7} \\ H^{(0)}_a \\ H^{(0)'a}
\end{array} \right).
\end{equation}

\noindent The vanishing of the lower-right 2-by-2 block is due to
the absence of a mass term of the form $M (\overline{{\bf 7}}_H^0
{\bf 7}_H^0)$ (or, equivalently, the vanishing of the VEV of
$Z^{\prime \prime}$). There is one zero eigenvalue of this matrix,
corresponding to a pair of massless multiplets with $Y/2 = \pm
\frac{1}{6}$ that gets eaten. All the other fields of this type get
superheavy mass.

There is finally a single pair of Higgs(ino) fields in $(3,1,
\frac{2}{3}) + conj.$, namely $H^{(1)a}$ and $H^{(-1)}_a$.  These
have no mass term, since there is no mass term of the form
$M(\overline{{\bf 7}}_H^{-1} {\bf 7}_H^1)$ (or, equivalently, the
VEV of $Z$ vanishes). These are goldstone modes that get eaten. The
mass matrix of the doublet Higgs(ino) fields that are in $(1,2,
\frac{1}{2}) + conj.$ is

\begin{equation}
W_{(1,2,1/2)} = \left( \begin{array}{c} H_{i67} \\ H_{i6} \\
H^{(1)}_i \\ H^{(1)i} \end{array} \right)^T \left(
\begin{array}{cccc} M_4 & 0 & 0 & 0 \\
0 & 0 & 0 & 0 \\
0 & 0 & M_1 & 0 \\
0 & 0 & 0 & 0 \end{array} \right) \left( \begin{array}{c} H^{i67}
\\ H^{i6} \\ H^{(-1)i} \\ H^{(-1)}_i \end{array} \right).
\end{equation}

\noindent There are two zero eigenvalues, corresponding to two pairs
of massless multiplets with $Y/2 = \pm \frac{1}{2}$. One of these
pairs is eaten, as we have seen above. This leads to a single
massless pair of doublets that are the doublet Higgs(ino) fields of
the MSSM.

Finally, the are Higgs(ino) fields in $(1,2,0)$. These have a mass
matrix

\begin{equation}
W_{(1,2,0)}= \left( \begin{array}{c} H^{i7} \\
H^{(0)i} \\ H^{(0)'i}
\end{array} \right)^T \left(
\begin{array}{ccc}
0 & \langle H^{(0)'}_7 \rangle & \langle H^{(0)}_7 \rangle\\
\langle H^{(0)'7} \rangle & 0 & 0 \\
\langle H^{(0) 7} \rangle & 0 & 0
\end{array} \right) \left( \begin{array}{c}
H_{i7} \\ H^{(0)}_7 \\ H^{(0)'}_i
\end{array} \right).
\end{equation}

\noindent here there is one zero eigenvalue, corresponding to the
pair of massless multiplets that gets eaten. All other fields of
this type get superheavy mass.

In sum, the only Higgs(ino) fields that do not get eaten or get
superheavy mass are the pair of doublets of the MSSM.  In other
words, there is successful doublet-triplet splitting.

We have shown that the missing partner mechanism can be implemented
in larger unitary gauge groups, by a straightforward generalization
of the well-known implementation in flipped $SU(5) \times U(1)$. The
details are more involved, due to the larger groups and the
correspondingly larger multiplets. But the essential idea is no more
complicated than in flipped $SU(5)$. The examples that have been
worked out in detauil here show that the generalization to unitary
groups larger than $SU(5)$ does not require complicated conditions
to be imposed on the superpotential or the structure of the theory.

\end{document}